\documentclass[11pt]{article}
\usepackage{times}
\usepackage{geometry}
\usepackage[utf8]{inputenc}
\usepackage{enumitem,amssymb}
\usepackage{ragged2e}
\newlist{thematic}{itemize}{8}
\setlist[thematic]{label=$\square$}
\usepackage{pifont}

\newcommand{\myvspace}{\vspace{0.1cm}}
\usepackage{enumitem}
\setlist[itemize]{leftmargin=*}
\usepackage{fancyhdr}
\usepackage[USenglish]{babel}
\usepackage[utf8]{inputenc}
\usepackage[T1]{fontenc}
\usepackage{epsfig}
\usepackage{wrapfig}
\usepackage{color}
\usepackage[vflt]{floatflt}
\usepackage{longtable}
\usepackage{caption}
\usepackage{jheppub}   
\usepackage[dvipsnames,svgnames,x11names]{xcolor}
\usepackage[all]{hypcap} 
\usepackage{amssymb,amsmath}
\usepackage{longtable}
\usepackage{amssymb}
\usepackage{amsmath, xspace}
\usepackage{graphics,graphicx} 
\usepackage{rotating}
\usepackage{color}
\usepackage{xcolor}
\usepackage{multirow}
\usepackage{subfigure}
\usepackage{sectsty}
\usepackage[outercaption]{sidecap}
\usepackage{eso-pic,graphicx}
\usepackage{tikz}

\usepackage{xcolor}
\definecolor{cobalt}{rgb}{0., 0.35, 0.56}
\usepackage{hyperref}

\AddToShipoutPictureBG{%
  \ifnum\value{page}=1
    \AtPageLowerLeft{%
      \includegraphics[width=\paperwidth]{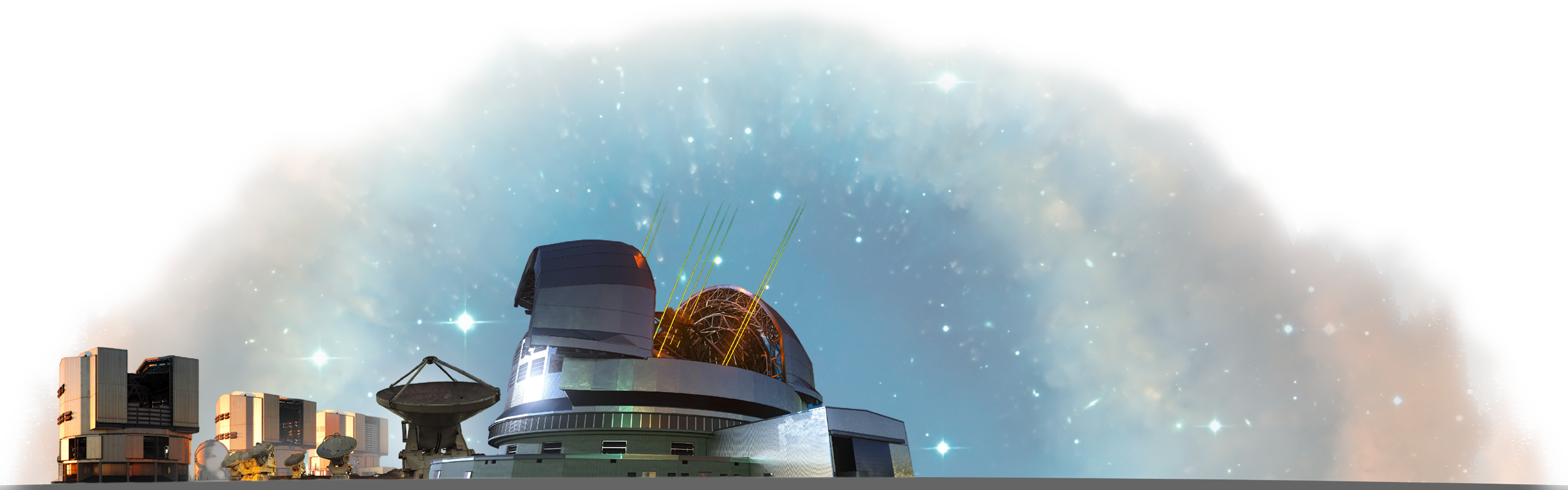}%
    }%
  \fi
}

\sidecaptionvpos{figure}{c}

\usepackage{enumitem}
\setlist[enumerate]{itemsep=0pt, parsep=0pt}

\definecolor{DarkGreen}{rgb}{0.0, 0.3, 0.0}
\definecolor{purple}{rgb}{0.5, 0.0, 0.5}
\definecolor{red}{rgb}{1, 0.0, 0.0}
\definecolor{green}{rgb}{0, 1.0, 0.0}

\def\3he{$^3{\rm He}$}

\def\lsim{\mathrel{\lower2.5pt\vbox{\lineskip=0pt\baselineskip=0pt
           \hbox{$<$}\hbox{$\sim$}}}}

\def\gsim{\mathrel{\lower2.5pt\vbox{\lineskip=0pt\baselineskip=0pt
           \hbox{$>$}\hbox{$\sim$}}}}

\begin{document}

\raggedright
\Large
ESO Expanding Horizons initiative 2025 \linebreak
Call for White Papers

\vspace{1.cm}

\raggedright
\huge
Transients as Determinants of Habitability 

\vspace{0.3cm}
\normalsize

\raggedright

\textbf{Scientific Question:} How might ESO’s next major telescope transform our understanding of stellar magnetic activity and its implications for exoplanet habitability in the era of the Habitable Worlds Observatory?

\vspace{0.3cm}

\normalsize

\bigskip

\textbf{Authors:} 
Fatemeh Zahra Majidi (fatemeh.majidi@inaf.it, INAF OACN, Italy); Katia Biazzo (INAF - Osservatorio Astronomico di Roma, Italy); Maria Tsantaki (INAF - Osservatorio Astrofisico di Arcetri); Amelia Bayo (ESO-Garching, Germany); Gra\v{z}ina Tautvai\v{s}ien\.{e} (Vilnius University, Lithuania); Valentin D. Ivanov (ESO-Garching, Germany); Germano Sacco (INAF-Osservatorio Astrofisico di Arcetri); Richard I. Anderson (EPFL, Switzerland); Avraham Binnenfeld (EPFL, Switzerland); David Montes 
(Universidad Complutense de Madrid and IPARCOS-UCM, Spain)

\myvspace

\textbf{Science Keywords:} 
young exoplanet hosts, habitability, stellar magnetic activity
\linebreak

\begin{figure}[!ht]
   \centering
   \includegraphics[width=0.5\linewidth]{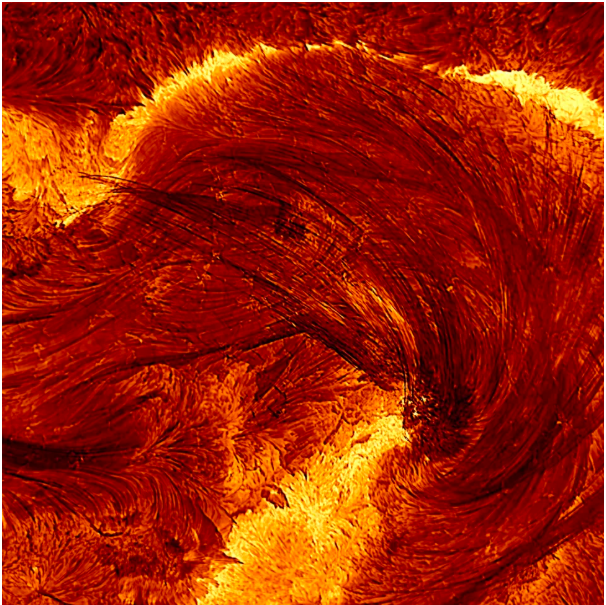}
   \caption*{\small Inouye Solar Telescope image of a solar flare on August 8, 2024. (Image credit: NSF/NSO/AURA, CC-BY)}
   \label{fig:placeholder}
\end{figure}

\clearpage
\textbf{\large Abstract:} Stellar magnetic activity, manifested through spots (faculae and flares), fundamentally shapes the exoplanets' environments. For low-mass stars in particular, where most habitable-zone planets reside, the variable magnetic phenomena can dominate atmospheric chemistry, surface radiation levels, long-term atmospheric escape, and ultimately habitability. However, physical characteristics of these transients (e.g. energy and temperature) and their spectra remain ill-constrained due to limitations in cadence and magnitude access of current spectroscopic facilities. A next-generation 12-m class ground-based observatory equipped with integral-field spectroscopy (IFS) and multi-object spectroscopy (MOS) at R$\sim$4,000 and $\sim$40,000 offers a transformational opportunity to characterize stellar activity in the time domain across large samples of exoplanet host stars. Such a facility would enable simultaneous monitoring of continuum variability, chromospheric and coronal line diagnostics, and particle-accelerated flare signatures, resolving the physics driving space weather and quantifying its impact on planetary atmospheres.

\vspace{5mm}
{\bf\Large 1. Science Drivers} 

The classical habitable zone framework assumes relatively stable insolation, yet the radiation and particle injection from stellar magnetic activity can dominate the atmospheric evolution of close-in terrestrial worlds. Active stars, particularly M dwarfs, exhibit frequent flares with energies spanning 10$^{30}$–10$^{35}$ erg, orders of magnitude higher than solar analogs on a per-unit-luminosity basis. These events generate impulsive UV/blue continuum enhancements, strong Balmer-series emission, and high-energy particle fluxes capable of altering ozone and NOx/HOx chemistry, driving atmospheric escape through enhanced ionization, increasing surface UV-C irradiation levels, modulating cloud formation and photochemistry, and producing prebiotic precursor molecules under specific conditions (Berger et\,al. 2024). \myvspace

The net impact is uncertain: flares may be hazardous or catalytic for life, depending on frequency, spectral energy distributions (SEDs), and contextual environmental factors (Rimmer et\,al. 2018; Tilley et\,al. 2019). To determine which regimes promote or suppress habitability, we require time-resolved spectroscopy that captures both the transient continuum slopes and the rich line forest associated with chromospheric heating, with sufficient cadence to track impulsive rise phases lasting seconds to minutes. \myvspace

Stellar magnetic activity can significantly influence the dynamics that may lead to the engulfment of planets. Stellar magnetic activity can lead to tidal forces that might cause the planet to lose orbital stability over time. This can lead to scenarios where a planet spirals inward and eventually gets engulfed by the star (Gehan 2025).  For stars with high magnetic activity, magnetic braking can slow down the rotation of the star. This can alter the stellar wind and the magnetic environment around the star, impacting the stability of nearby planets (Matt \& Pudritz 2005). Magnetic interactions can also affect planetary migration patterns, potentially leading to collisions or absorption by the star if a planet's orbit decays sufficiently (Nelson \& Papaloizou 2004). Further detailed studies are necessary to fully understand the mechanisms involved. \myvspace

Space-based missions (Kepler, TESS) have provided broadband photometric flare catalogs, yet they lack spectroscopic diagnostics needed to determine temperatures, densities, energies, and particle fluxes. Current ground-based facilities offer limited sample size, sensitivity, or cadence, particularly for fainter M dwarfs and for high-resolution spectroscopic coverage of large stellar populations. A 12-m class facility dedicated to time-domain astronomy and monitoring large areas of the sky can provide order-of-magnitude improvements in temporal coverage, sensitivity, and diagnostic richness like never before. \myvspace

The combination of aperture, multiplexing, and spectral resolution is uniquely suited to deliver flare \textbf{I)} SEDs from 350–2000\,nm at high signal-to-noise, revealing the Balmer and Paschen continua, blackbody components, and radiative backwarming; \textbf{II)} Velocity-resolved chromospheric line diagnostics (e.g., H$\alpha$, Ca II H$\&$K, He I 1083\,nm) with R$\sim$40,000 for measuring mass motions, turbulence, and reconnection-driven evaporation; \textbf{III)} Population-level flare rates and energies across hundreds of exoplanet hosts; \textbf{IV)} Spot and faculae evolution through line-profile tomography, enabling improved stellar surface mapping and correction of stellar contamination in transmission spectroscopy. \textit{These multi-scale time-domain datasets will provide the first unified empirical framework linking stellar activity physics to exoplanet atmospheric stability and surface habitability.}

\vspace{3mm}
{\bf\Large 2. Synergies with other facilities}

A 12-m class ground-based facility with high-cadence operational for multiple years, equipped with IFS and MOS provides essential time-domain and spectroscopic context for the next generation of exoplanet missions. These missions increasingly confront the limitations imposed by stellar activity(spots, faculae, and especially flares) which imprint spectral and photometric signatures at or above the precision levels required for atmospheric characterization. Continuous, multi-wavelength, high-resolution monitoring from the ground enables calibration, validation, and enhancement of space-based exoplanet science across several key missions. The synergy between different facilities addresses the transformative concepts outlined below:
    
\textbf{1. Transforming Atmospheric Characterization through Stellar Context}

Atmospheric measurements from Ariel, JWST, Roman, and ultimately HWO demand exquisite correction for stellar contamination. Without contemporaneous knowledge of flare activity, spot/faculae evolution, or chromospheric state, transmission and emission spectra can be fundamentally misinterpreted. \textit{This synergy elevates space-based atmospheric retrievals from descriptive to physically interpretable, ensuring reliable abundance, climate, and even biosignature assessments (Rackham \& de Wit 2024).}

\textbf{2. Unlocking Population-Level Insights into Stellar Activity and Planetary Environments}

Wide-field missions such as Euclid and LSST/Rubin will generate enormous catalogs of variable stars, flares, transiting planets, and microlensing hosts. Yet photometry alone cannot identify the physical drivers of variability nor the extremity of stellar environments that shape planetary atmospheres. The 12-m facility will provide massive-scale MOS spectroscopy to characterize the magnetic activity and their associated physical features/processes of the very populations these missions discover. This will in turn, enable linking population variability maps from Euclid and LSST with spectroscopic diagnostics that reveal the underlying magnetic architectures.
\textit{These synergies allow demographic studies of space weather in the Galactic context, which no mission can achieve alone.}

\textbf{3. Delivering Predictive Stellar Space-Weather Models for Habitable Planet Searches}

The search for life (especially with HWO, but also with JWST and Ariel) requires understanding whether exoplanet atmospheres can persist under real stellar irradiation and particle conditions (Berger et\,al. 2024); e.g. 
assessing exoplanet habitability requires direct spectroscopic characterization of the stellar environments that shape atmospheric survival. Existing facilities lack the time-resolution, wavelength coverage, and spectral fidelity needed to track how stellar radiation fields and particle-accelerating events evolve in real time to build predictive models of space weather.

A 12-m spectroscopic facility would provide \textbf{I)} continuous, high-resolution, multi-epoch spectroscopy capable of tracing flare energetics, magnetic-field reconfigurations, and long-term activity cycles; \textbf{II)} spectral diagnostics that quantify the full distribution of energetic, particle-producing flares and their atmospheric impact regimes; and \textbf{III)} empirically calibrated spectroscopic space-weather models that feed directly into HWO target selection, identifying stars whose environments are most likely to support stable, long-lived, potentially life-bearing atmospheres.

Spectroscopic IFS would substantially enhance stellar space-weather and habitability studies by providing time-resolved spectral cubes that capture the full spatial and spectral evolution of magnetic activity. IFS enables mapping of flare kernels and active regions, tracking temperature, density, and velocity structures during eruptions, and simultaneously monitoring key diagnostic lines that reveal heating, particle acceleration, and potential CME signatures. Its broad, multi-line coverage and rapid cadence deliver far richer constraints on stellar irradiation and particle environments than traditional spectroscopy, while also helping disentangle stellar variability from planetary signals and identify possible star–planet magnetic interactions. Together, these capabilities allow IFS to produce the detailed, empirical stellar-environment models needed to predict atmospheric survival and guide target selection for missions like the Habitable Worlds Observatory.

\vspace{3mm}
{\bf\Large 3. Key Questions}

Here are the major open questions that the next big telescope developed by ESO will address through conducting a decadal spectroscopic survey of young, active exoplanet hosts:

{\bf(I)} How do flare and spot activity translate into atmospheric escape and photochemistry on exoplanets?

{\bf(II)} How do chromospheric and coronal plasma dynamics evolve in real time during transient magnetic events, and what do these dynamics reveal about mass motions, magnetic topology, and particle injection relevant for planetary atmospheric loss?

{\bf(III)} What are the fundamental physical mechanisms that trigger and power the most energetic stellar flares across various stellar spectral types, and how do they reshape exoplanetary environments on timescales from seconds to gigayears?

{\bf(IV)} How do spots and faculae evolve, and how do they bias exoplanet measurements? 

{\bf(V)} How do these events influence protoplanetary chemistry, ionization structure, and the early atmospheric evolution of close-in forming planets?

{\bf(VI)} How does the full-time-resolved SEDs of stellar flares (homogeneously from the optical to the NIR regime) control the photochemical trajectories and long-term habitability of exoplanet atmospheres? Can life survive or \textit{even originate} within the intense, spectrally complex flare environment of common M dwarfs?

{\bf(VII)} Under what conditions does stellar magnetic activity transition from atmospheric destroyer to chemical catalyst for prebiotic molecules, and which stars represent the most promising targets for biosignature searches?

\vspace{5mm}
\textbf{References:}
Berger et\,al. 2024, MNRAS, 532, 4436;
Rimmer, et\,al. 2018, Sci.Adv., 4, 3302;
Tilley, et\,al. 2019, Astrobio., 19, 64;
Gehan 2025, A\&A, in press [arXiv:2510.27295];
Matt \& Pudritz 2005, MNRAS, 356, 167;
Nelson \& Papaloizou 2004, MNRAS, 350, 849;
Rackham \& de Wit, J. 2024, AJ, 168, 82.

\vspace{5mm}
{\bf\Large Technical Requirements}

Ideally, to meet our science goals the spectroscopic facility will meet the following requirements:

{\bf(I)} Aperture and Sensitivity: 12-m class to achieve S/N $\geq$ 100 per resolution element at V$\sim$14$^m$ in $\leq$60 s exposures which enables real-time capture of flare rise phases and faint M-dwarf hosts.

{\bf(II)} Spectral Resolution: A) Low-res (R$\sim$4,000) IFS mode for capturing SED evolution, broad features, and continuum slopes. B) Low- and high-res (R$\sim$40,000) MOS modes for chromospheric lines, velocity fields, and detailed line diagnostics.

{\bf(III)} Wavelength Coverage: 350–2000\,nm, with simultaneous optical + NIR channels preferred. Must include H$\alpha$, H$\beta$, Ca II H$\&$K, Ca II IR triplet, He I 5876 $\AA \&$ 1083\,nm, Fe II multiplets, continuum windows around Balmer/Paschen jumps.

{\bf(IV)} Time Resolution and Cadence: I) 1–10 s cadence for bright targets in IFS mode to resolve impulsive flare phases.
II) $\leq$1 min cadence for MOS monitoring of large target lists. 
III) Continuous coverage for $\geq$6 hr to capture flare sequences and rotation.

{\bf(V)} Field of View and Multiplexing: 
I) MOS: $\geq$50 simultaneous targets to build population-level frequency–energy distributions. II) IFS: field of view sufficient to include sky calibrators and maximize flare-capture probability for known hosts.

{\bf(VI)} Calibration and Stability: Rapid readout with minimal dead time for true time-domain operation.

\end{document}